\documentclass[12pt]{article}

\usepackage{amsmath,amssymb,amsthm}
\usepackage{amsmath,amssymb}

\newcommand{\n}{\noindent}

\newcommand{\ka}{\kappa}
\newcommand{\ed}{\end{document}}
\newcommand{\be}{\begin{equation}}
\newcommand{\ee}{\end{equation}}

\begin{document}
\begin{center}
\large{\textbf{\textbf{Towards a Discrete Spacetime}}}\\
\end{center}
\begin{center}
Sudipta Das\footnote{E-mail: sudipta.das\_r@isical.ac.in} and
Subir Ghosh\footnote{E-mail: sghosh@isical.ac.in} 
\\
Physics and Applied Mathematics Unit, Indian Statistical
Institute\\
203 B. T. Road, Kolkata 700108, India \\
\end{center}\vspace{1cm}

\begin{center}
{\textbf{Abstract}}
\end{center}

\n  A formalism is proposed to generate (the first step of) a discrete
spacetime: spacetime with an inbuilt
length scale. We follow the celebrated Landau theory of liquid - solid phase
transition induced by 
Spontaneous Symmetry Breaking by a condensate whose Fourier transform has
support at a {\it{non-zero}} momentum.
The latter requirement is essential for breaking the translation symmetry. This,
in turn, compels us to generalize
Einstein action to higher derivative terms. 

\vskip 1cm

Is there a fundamental way to introduce a (short distance) length
scale, {\it{eg}}. the Planck scale, in Einstein's General
Relativity (GR)? It can then serve as an effective theory of  Quantum
Gravity. So far the attempts have been along the lines of
Non-Commutative (NC) geometry framework \cite{ncgr} where one
exploits Seiberg Witten map \cite{sw} to incorporate NC
corrections on GR. In the present paper we follow a completely
different route taking cue from Condensed Matter Physics: the
Landau theory of liquid-solid phase transition \cite{lan}.

In a
series of pioneering works by Alexander and Mctague \cite{alex} it
was explicitly demonstrated how one is able to construct a
{\it{discrete lattice}}, (where translation and rotation
symmetries are lost), from a liquid phase, (where the symmetries
are intact), through Spontaneous Symmetry Breaking (SSB) with a
crucial additional requirement: the (Fourier transform of)
non-zero VEV of condensate  that minimizes the free
energy must have
support at a non-vanishing momentum. The difference between the (inhomogeneous)
density for solid
and (constant) density of liquid acts as  the order parameter. The free energy
is expanded around the
higher symmetry liquid phase and it should be mentioned that to form a proper
crystal lattice one needs terms,  at least of third and fourth order in the
order parameter, in the expansion.   This idea was later applied by 
Rabinovici et.al.
\cite{rab} in string theory compactification. The beauty of the
formalism lies in its universality and independence of details of
the specific model in question. More explicitly \cite{alex,rab,lan}, consider
the density difference,
 $\rho (\vec x)=\rho _{solid}(\vec x)-\rho _{liquid}$,
between (lower symmetry) solid phase $\rho _{solid}(\vec x)$ and symmetric
liquid phase with constant $\rho_{liquid}$ and
take its Fourier decomposition, $\rho (\vec x)= \sum \rho (\vec p)exp(i\vec
p.\vec x) +h.c.$. Now, from the first non-trivial
quadratic term in the effective action $L \sim \int d\vec p_1d\vec p_2\rho(
p_1)\rho( p_2)A(\mid \vec p_1\mid ^2)
\delta (\vec p_1-\vec p_2)$, one needs to ascertain the value of $\mid \vec p_1
\mid $ that will minimize $A(\mid \vec p_1\mid ^2)$.
If it is different from zero, $\rho (\vec x)$ breaks translation invariance, but
still rotation invariance is intact
since $\mid \vec p_1\mid ^2$ is involved. Reality of $\rho (\vec x)$ demands
$\rho (\vec p)=\rho (-\vec p)$ and one gets
$\rho (\vec x)\sim cos (\vec q .\vec x)$. The third and fourth order terms in
$L$
introduces more structure to
$\rho (\vec x)$ that breaks rotational invariance as well \cite{alex,rab}. The
order parameter can have tensorial structure as well. In our case where the
``solidification'' of 
spacetime is involved, the role of order parameter will be played by the
linearized part 
of the metric $g_{\mu\nu}\approx \eta_{\mu\nu}+h_{\mu\nu}$.

Our longterm goal is to generate a spacetime lattice through SSB in the
above way. However, in the present article our target is more modest:
as a first step towards this objective, we intend to introduce a
short distance scale ($\sim$ Planck length) in the GR action that
will break the translation invariance, without affecting the
rotational symmetry. Indeed, the extension of GR to NC spacetimes
\cite{ncgr} also aims to achieve that but, compared to this ad-hoc
procedure, our approach is much more basic and intuitive.   We
introduce  the deformation directly in the metric fluctuation over
the condensate, that plays the role of a (tensorial) order
parameter in the continuum-discrete spacetime phase transition. At the
same time it is very interesting and encouraging that our proposed
form of the metric bears important similarities with the NC
extended metric of \cite{ncgr}.

 There is an elegant way of
interpreting the masslessness of photon: they are the Goldstone
bosons induced by SSB of diffeomorphism invariance \cite{ssb}. We,
in particular, extend the work of Kraus and Tomboulis \cite{kr} by
taking in to account higher derivative terms in the action,
$R^{\mu\nu}R_{\mu\nu}$ and $(\eta ^{\mu\nu}R_{\mu\nu})^2$.  
Apart from the conventional $(\eta ^{\mu\nu}R_{\mu\nu})$-term, the above are
needed to ensure that the kinetic
term is minimized for a non-zero momentum. 
All the subtle
arguments of \cite{kr}, regarding the degrees of freedom count
after SSB that yields the massless physical graviton (to lowest
order) remains applicable in our model as well since the higher
order terms respect the same symmetries as GR \cite{ste}. The only
possible ambiguity can arise from the fact that the higher
derivative massive spin two states can bring in negative energy
states \cite{ste}. However, once again, falling back to the argument of
\cite{kr}, these massive modes possibly do not affect the low
energy physics, at least for length scales large compared to Planck length
\cite{ste} (the parameter
introduced by us in the present work).

We start with the higher derivative lagrangian $L$ \be L = \int
d^4 x {\cal{L}} = \int d^4 x [\sqrt{g} \left(R + \alpha R^2 + \beta
R_{\mu \nu} R^{\mu \nu} \right) +V] \label{lag} \ee where   $g_{\mu
\nu}, R_{\mu \nu},R$  are respectively  the  metric tensor, Ricci
tensor and Ricci scalar.  $\alpha$ and $\beta$ are numerical
parameters. $V$ is some suitable symmetry breaking potential that
can appear {\it{e.g.}} from effective one-loop matter coupling
\cite{kr}.

Throughout we will work in  weak field approximation with $ g_{\mu
\nu} = \eta_{\mu \nu} + h_{\mu \nu}~~,~~|h_{\mu \nu}| \ll
1,~~\eta_{\mu \nu} = diag(-1, 1, 1,1) $. Considering  terms up to
second order in $h$ (i.e. neglecting terms of $O(h^3)$ and higher)
${\cal{L}}$ in (\ref{lag}) reduces to
 \be {\cal{L}} = \frac{1}{4}
h \Box h - \frac{1}{4} h^{\rho \sigma} \Box h_{\rho \sigma} +
\frac{1}{2} h^{\rho \sigma} \partial_{\mu}
\partial_{\rho} h^{\mu}_{\sigma} - \frac{1}{2} h^{\mu \nu}
\partial_{\mu} \partial_{\nu} h - \alpha h \Box^2 h + 2 \alpha
h^{\lambda \mu} \partial_{\mu}
\partial_{\lambda} \Box h - \alpha h^{\lambda \mu} \partial_{\mu}
\partial_{\lambda} \partial_{\rho} \partial_{\sigma} h^{\rho
\sigma} $$$$ - \frac{\beta}{4} h \Box^2 h + \frac{\beta}{2}
h^{\lambda \mu} \partial_{\mu} \partial{\lambda} \Box h -
\frac{\beta}{2} h^{\lambda \mu} \partial_{\mu} \partial_{\lambda}
\partial_{\rho} \partial_{\sigma} h^{\rho \sigma} +
\frac{\beta}{2} h^{\mu \ka} \partial_{\lambda} \partial_{\mu} \Box
h^{\lambda}_{\ka} - \frac{\beta}{4} h^{\mu \ka} \Box^2 h_{\mu \ka}
+V . \label{lagh} \ee We have used the notation $h = \eta^{\mu
\nu} h_{\mu \nu},\Box = \partial_{\mu}
\partial^{\mu}$. Furthermore, if we put $\alpha = - \frac{\beta}{2}$ in
(\ref{lagh}), it takes
the following form \be {\cal{L}} = \frac{1}{4} h \Box (1 - 2
\alpha \Box) h - \frac{1}{2}h^{\mu \nu} \partial_{\mu}
\partial_{\nu} (1 - 2 \alpha \Box) h - \frac{1}{4} h^{\rho \sigma}
\Box (1 - 2 \alpha \Box) h_{\rho \sigma} + \frac{1}{2} h^{\rho
\sigma} \partial_{\mu} \partial_{\rho} (1 - 2 \alpha \Box)
h^{\mu}_{\sigma}+V. \label{lagwg} \ee 
The structure is essentially same as the Einstein gravity only with the operator
$\sim -2\alpha \Box $ appended.

One can still  use the
harmonic gauge $
\partial_{\mu} h^{\mu \nu} = \frac{1}{2}
\partial^{\nu} h, \label{hg} $ to get \cite{ste}
\be {\cal{L}} = \frac{1}{8} h (1 - 2 \alpha \Box)\Box h -
\frac{1}{4} h^{\rho \sigma} (1 - 2 \alpha \Box)\Box h_{\rho
\sigma}+ V. \label{laghg} \ee But we will not consider the last
form \cite{kr}.

To consider SSB effects, let us consider fluctuations
${\tilde{h}}_{\mu \nu}$ above the condensate:
 \be g_{\mu \nu} = \eta_{\mu \nu} + h_{\mu \nu} = \eta_{\mu \nu}
+ <h_{\mu \nu}> cos (q.x) + {\tilde{h}}_{\mu \nu}(x), \label{metric}
\ee where $<~~>$ represents vacuum expectation value and $q_{\mu}$
is a non-zero vector with $q^2=-1/(4\alpha )$. This is the crucial
extension of \cite{kr} that comes from minimization of the kinetic
energy part in (\ref{laghg}), thanks to the higher derivative
terms. The specific form, $cos (q.x)$ comes from symmetry
arguments while treating ${\tilde{h}}_{\mu \nu}$ as the order
parameter (as explained briefly, for details see \cite{alex,rab}).  
The normalization is fixed from minimizing the kinetic 
part of the action, $-k^2(1+2\alpha k^2)$ in (\ref{l}) below and identifying
$k^\mu
\equiv q^\mu $. Since
we would like to interpret $q^\mu $ as the Planck momentum this requires $\alpha
$ to be small indicating
that effect of the higher derivative terms in the action are small. It is now
clear that to make a full fledged 
spacetime ``crystal'' one requires third and fourth terms in $h_{\mu\nu }$, (as
explained in 
\cite{alex,rab} for liquid-solid transition), in
the action which will considerably
complicate the model.

In terms of ${\tilde{h}}_{\mu \nu}$ (\ref{metric})  ${\cal{L}}$
becomes
 \be {\cal{L}} = [\frac{1}{4}
\eta^{\alpha \beta} \eta^{\lambda \sigma} {\tilde{h}}_{\alpha
\beta} \Box (1 - 2 \alpha \Box){\tilde{h}}_{\lambda \sigma}-
\frac{1}{2} \eta^{\mu \sigma} \eta^{\nu \lambda}
{\tilde{h}}_{\lambda \sigma} \eta^{\alpha \beta} \partial_{\mu}
\partial_{\nu} (1 - 2 \alpha \Box){\tilde{h}}_{\alpha \beta} $$$$
 -
{\frac{1}{4}} \eta^{\mu \sigma} \eta^{\nu \lambda}
{\tilde{h}}_{\lambda \sigma} \Box (1 - 2 \alpha
\Box){\tilde{h}}_{\mu \nu}+ \frac{1}{2} \eta^{\rho \alpha}
\eta^{\sigma \beta} {\tilde{h}}_{\alpha \beta} \eta^{\mu \lambda}
\partial_{\mu} \partial_{\rho} (1 - 2 \alpha
\Box){\tilde{h}}_{\lambda \sigma} ] $$$$
+cos (q.x)[{\frac{1}{4}} \eta^{\alpha \beta}
\eta^{\lambda \sigma} <h_{\alpha \beta}>  \Box (1 - 2
\alpha \Box){\tilde{h}}_{\lambda \sigma} $$$$
 -\frac{1}{2} \eta^{\mu \sigma} \eta^{\nu \lambda} <h_{\lambda
\sigma}>  \eta^{\alpha \beta} \partial_{\mu}
\partial_{\nu} (1 - 2 \alpha \Box){\tilde{h}}_{\alpha \beta}$$$$
- {\frac{1}{4}} \eta^{\mu \sigma} \eta^{\nu \lambda} <h_{\lambda
\sigma}>  \Box (1 - 2 \alpha \Box){\tilde{h}}_{\mu \nu}{\frac{1}{2}} \eta^{\rho
\alpha}
\eta^{\sigma \beta} <h_{\alpha \beta}>  \eta^{\mu
\lambda} \partial_{\mu} \partial_{\rho} (1 - 2 \alpha
\Box){\tilde{h}}_{\lambda \sigma} $$$$
 - {\frac{1}{4}} \eta^{\alpha
\beta} {\tilde{h}}_{\alpha \beta} q^2 (1 + 2 \alpha q^2)
<h_{\lambda \sigma}> \eta^{\lambda \sigma} $$$$
 + \frac{1}{2}
\eta^{\mu \sigma} \eta^{\nu \lambda} {\tilde{h}}_{\lambda \sigma}
q_{\mu} q_{\nu} (1 + 2 \alpha q^2) \eta^{\alpha \beta} <h_{\alpha
\beta}>  $$$$ + {\frac{1}{4}} \eta^{\mu \sigma} \eta^{\nu
\lambda} {\tilde{h}}_{\lambda \sigma} q^2 (1 + 2 \alpha q^2)
<h_{\mu \nu}>  - {\frac{1}{2}} \eta^{\rho \alpha}
\eta^{\sigma \beta} {\tilde{h}}_{\alpha \beta} q_{\mu} q_{\rho}
\eta^{\mu \lambda} (1 + 2 \alpha q^2) <h_{\lambda \sigma}> ] +V. \label{l} \ee
We have dropped the constant terms in
(\ref{l}).

Varying (\ref{l}) with respect to ${\tilde{h}}_{\alpha \beta}$ we
obtain the following equation of motion \be \frac{1}{2} (1 - 2
\alpha \Box)[\eta^{\alpha \beta} (\Box \tilde h -
\partial^{\mu} \partial^{\nu} {\tilde{h}}_{\mu \nu}) -
\partial^{\alpha} \partial^{\beta} \tilde h - \Box {\tilde{h}}^{\alpha \beta} +
\partial^{\alpha} \partial^{\mu} {\tilde{h}}^{\beta}_{\mu} +
\partial^{\beta} \partial^{\mu} {\tilde{h}}^{\alpha}_{\mu}] $$$$ +
\frac{1}{2} (1 + 2 \alpha q^2) cos (q.x) [\eta^{\alpha \beta}(-q^2
<h> + q_{\mu} q_{\nu} <h^{\mu \nu}>) $$$$ + q^{\alpha} q^{\beta}
<h> + q^2 <h^{\alpha \beta}> - q^{\alpha} q_{\mu} <h^{\mu \beta}>
- q^{\beta} q_{\mu} <h^{\mu \alpha}>] = 0. \label{em} \ee We will
not consider the effects of the potential term any further since
it will lead to mass terms and higher order interaction terms
\cite{kr}.

Notice that the second set of terms in (\ref{em}) is independent
of the field ${\tilde{h}}_{\mu \nu}$ and hence can be treated as a
source term, (of $O({\tilde h} ^0$), in order to solve (\ref{em}) for
${\tilde{h}}_{\mu
\nu}$. The source is symmetric and conserved but is not of the perfect fluid
form.  But to be a realistic source it should satisfy, at the least, the Weak
Energy Condition
 $t_\mu t_\nu \tilde {h}^{\mu\nu}\geq 0$, for arbitrary time-like $t_\mu $ (see
{\it{eg.}} \cite{car}). In the present case
this will lead to restricting the numerical values of $<h^{\mu\nu}>$ which is
perfectly justified. Also it 
might be interesting to study the properties of this source as a relativistic
imperfect fluid (see {\it{eg.}} \cite{wein}).

The equation is trivially solved by \be \tilde {h}_{\alpha\beta }
= cos (q.x) <h_{\alpha \beta}> .\label{h} \ee Thus in the weak
field limit we obtain \be
g_{\mu\nu}=\eta_{\mu\nu}+cos(q.x)<h_{\mu\nu}> .\label{g} \ee

It is
now straightforward to construct the Ricci tensor and scalar in an
explicit way:
 $$ R_{\mu\nu
}=-\frac{1}{2}(q_\sigma q_\nu <h^{\sigma } _\mu >+q_\sigma q_\mu
<h^{\sigma }_\nu >-q_\mu q_\nu <h> -q^2 <h_{\mu\nu}>)cos(q.x),$$ \be
R=\eta^{\mu\nu}R_{\mu\nu}=-(q_\mu q_\nu
<h^{\mu\nu}>-q^2<h>)cos(q.x). \label{rr} \ee 
Clearly the condensate induces an effective curvature. Depending on the choice
of parameters as well as the 
physically relevant form of curvature, it is  possible to put restrictions on
$cos(q.x) $. This can be translated to
an effective minimum length scale.

Our framework, though
considered from an entirely different perspective, compares
favorably  with  the NC-extended metric \cite{ncgr} in some basic features, provided one relates $\alpha $
with NC parameter.\\
I) In NC gravity NC effects generate a source term. In SSB scenario
the condensate induces a source, in the form of an exotic fluid.\\
II) It has been established in various frameworks and diverse
types of noncommtative structures that the NC corrections start
from quadratic order of the NC parameter. There are generically no
NC correction in linear order in the NC parameter. This is true in
the SSB formalism as well since $cos(q.x)\approx 1+(q.x)^2 +...$. However it is
probably
not justified to treat $(q.x)$ as an expasion parameter unless $x$ is very
small.

To conclude, we have shown a way in which  higher derivative terms, extending the Einstein action, can be 
used to construct a discrete spacetime where translation and rotation symmetries can be broken. In the present work we
have studied the first step towards this goal: introduction of a length scale in the metric. This modification
generates an imperfect fluid like source term that is reminiscent of noncommutative extensions of gravity 
models \cite{ncgr}. One has to consider in more detail the important issue of negative energy that plagues
higher derivative gravity theories since, (although ``numerically'' small), it plays an essential role 
in our scheme. The properties of the exotic fluid that appears in our model and it effects on the dynamics of
test particles need to be discussed.

\vskip .3cm
{\it{Acknowledgments}}: It is a pleasure to thank Professor Chetan Nayek and
Professor Eliezer Rabinovici for
suggestions and Professor Robertus Potting and Professor Partha Mitra for
correspondence and discussions.

\end{document}